\documentclass[twocolumn,showpacs,preprintnumbers,amsmath,amssymb]{revtex4}
\usepackage{graphicx}
\usepackage{dcolumn}
\usepackage{bm}

\begin{document}

\title{Light quark mass ratio from Dalitz plot of
$\eta \rightarrow \pi^+\pi^-\pi^0$ decay}
\author{B.V. Martemyanov}
\author{V.S.  Sopov}
\affiliation{Institute for Theoretical and Experimental
Physics, B. Cheremushkinskaya 25, 117259 Moscow, Russia}

\begin{abstract}
High statistics Dalitz-plot distribution of $\eta \rightarrow \pi^+\pi^-\pi^0$ decay
obtained recently by KLOE collaboration~\cite{kloe} is fitted to the results of corresponding
theoretical calculations in Chiral Perturbation Theory (ChPT) with unitarity corrections
taken into account. 
 The quark mass ratio  $Q = \sqrt{(m^2_s - (m_d + m_u)^2/4)/(m^2_d - m^2_u)} $
can be otained from this analysis. We get $Q= 22.8\pm 0.4$
which differs from the  value $Q_{DT} = 24.2$ that follows from
Dashen's theorem and agrees with recently calculated electromagnetic kaon mass
difference.

\end{abstract}

\maketitle

The possibility to extract light quark mass difference from $\eta \rightarrow \pi^+\pi^-\pi^0$ decay
is known for a long time~\cite{L}. In ChPT the decay width $\Gamma$ depends on quark mass ratios and
theoretically calculable factor $\bar{\Gamma}$~\cite{GL}:

\begin{equation}
\Gamma = \left(\frac{Q_{DT}}{Q}\right)^4 {\bar \Gamma}~,
\label{gamma1}
\end{equation}
where
\begin{equation}
Q^{-2} = \frac{m^2_d - m^2_u}{m^2_s - {\hat m}^2}~,~~~{\hat m}=
\frac{m_d + m_u}{2}~,
\label{gamma2}
\end{equation}
$m_u , m_d , m_s$ - are up, down and strange quark masses,
\begin{equation}
Q^{-2}_{DT} = \frac{\left(\left( m^2_{K^0}-m^2_{K^+}\right)
- \left( m^2_{\pi^0}-m^2_{\pi^+}\right)\right)  m^2_{\pi^0}}
{\left( m^2_K - m^2_{\pi^0}\right) m^2_K} = (24.2)^{-2}~,
\label{gamma3}
\end{equation}
with $m^2_K =\left( m^2_{K^+}+m^2_{K^0}-m^2_{\pi^+}+m^2_{\pi^0}\right)/2$.

Note that $Q_{DT} = Q$ if electromagnetic mass differences of kaons and
pions are equal to each other as Dashen's theorem states~ \cite{Dashen}.
Experimental (Particle Data Group) value $\Gamma = 291\pm 21 ~{\rm eV}~$~\cite{PDG} is far from
one-loop ChPT value ${\bar \Gamma} = 167 \pm 50 ~{\rm eV}~$~\cite{GL} and
from the values  ${\bar \Gamma} = 209 \pm 20 ~{\rm eV}~$ ~\cite{Kambor} and
 ${\bar \Gamma} = 219 \pm 22 ~{\rm eV}~$ ~\cite{Anis}
 obtained with higher order
corrections taken into account by dispersion method.
In ~\cite{Kambor} the subtraction polinomial was taken from the decomposition
of one-loop order amplitude and had therefore uncertainties connected to
higher orders ChPT corrections. These uncertainties were further fixed~\cite{MS} by
fitting the experimental data~\cite{gormley} on Dalitz-plot distribution in the decay considered:
${\bar \Gamma} = 213^{+3}_{-12} ~{\rm eV}~$. There was a conjecture in~\cite{MS} that new
experimental data on Dalitz-plot distribution will give slightly different value of
${\bar \Gamma}$. Now these new experimental data (contradicting to the old ones~\cite{gormley})
are available~\cite{kloe}. In what follows we will use them to get new value of
${\bar \Gamma}$ and new value of quark mass ratio $Q$, as a consequence.

We use the method of work~\cite{MS} and remind it here for completeness.
In order to simulate the experimental Dalitz-plot distribution
we take it in a form
\begin{equation}
1+ay+by^2+cy^3+dx^2
\end{equation}
with $a=-1.075\pm 0.008$, $b=0.118\pm 0.009$, $c=0.13\pm 0.02$, $d=0.049\pm 0.008$~\cite{kloe}
and $y,x$ defined in a standart way
$$y=\frac{3T_0}{Q}-1~,~x=\frac{\sqrt{3}}{Q}\left(T_+-T_-\right)~,~
Q=T_++T_-+T_0~,$$
$T_+,T_-,T_0$ are the kinetic energies of pions in the rest frame
 $\eta \rightarrow \pi^+\pi^-\pi^0$ decay. We divide the Dalitz plot in $10\times 10$
 bins ($x\times y$)  that have equal number of events for the distribution considered.
 Then the number of events in each bin ($n$) is simulated by Gaussian distribution
 with variance equal to n. We used $n= 10 000$ to get the full statistics $N = 100 n =
1 000 000$ like in the experiment~\cite{kloe}. From theoretical point of view the amplitude
of $\eta \rightarrow \pi^+\pi^-\pi^0$ decay have an approximate solution from Eq.(5.28) of ~\cite{Kambor}.
It contains the subtraction polinomial
\begin{equation}
P(s) = \alpha +\beta s_a + \gamma s_a^2 + \delta (s_b-s_c)^2~,
\end{equation}
where $s_a, s_b$ and $s_c$ are invariant masses squared of $\pi^+\pi^-,
\pi^+\pi^0$ and $\pi^-\pi^0$ pairs, respectively. For the values of parameters
$\alpha, \beta, \gamma$ and $\delta$ within the regions
\begin{eqnarray}
&\alpha = -1.28 \pm 0.14,~&\beta = 21.81 \pm 1.52 ~{\rm GeV}^{-2}\nonumber \\
&\gamma = 4.09 \pm 3.18 ~{\rm GeV}^{-4},~&\delta = 4.19 \pm 1.08~{\rm GeV}^{-4}
\label{allowed} 
\end{eqnarray}
(the case of zero subtraction points~\cite{Kambor}) the ``Minuit'' fit
of above simulated experimental Dalitz-plot distribution has terminated on
the values
\begin{eqnarray}
&\alpha_0 = -1.14,~&\beta_0 = 23.33 ~{\rm GeV}^{-2}\nonumber \\
&\gamma_0 = 1.03q
 ~{\rm GeV}^{-4},~&\delta_0 = 5.27 ~{\rm GeV}^{-4} 
\end{eqnarray}
with $\chi^2/N{d.o.f.} = 152/(100-4)$. 

Three  from four parameters are at the boundary of allowed region (\ref{allowed}).
This probably means that the guess~\cite{Kambor} of the size of allowed region
should be changed.
Equally possible is the fit with the scaled values of parameters
$\alpha, \beta, \gamma$ and $\delta$ because the normalization factor of the 
amplitude is not defined by the Dalitz-plot distribution. In our case the scaling
of parameters $\alpha, \beta, \gamma$ and $\delta$ puts them outside the
allowed region (\ref{allowed}) and no freedom 
 in the scaling (no error in ${\bar \Gamma}$) is possible. This way we get
${\bar \Gamma} = 229~ ~{\rm eV}$ what corresponds according to eqs. (\ref{gamma1})-(\ref{gamma3})
to the light quark mass ratio $Q = 22.8\pm 0.4$. The errors here are due to the
errors in the experimental value of the width $\Gamma$. So, we conclude, high statistics
Dalitz-plot distribution gives the value of light quark mass ratio $Q$ slightly
lower than that from the assumption
of equality of kaon and pion electromagnetic mass differences ($Q_{DT} = 24.2$).
This is in agreement with calculations of electromagnetic mass differences
for pions and kaons~\cite{Donoghue,Bijnens} 
 which find large violations to Dashen's theorem ($Q = 22.0\pm 0.6$).
 Our result  agrees also
very well with that of works ~\cite{Kambor}   and~\cite{Anis},
where the values $Q=22.4\pm 0.9$ and $Q= 22.7\pm 0.8$ were obtained,
correspondingly.

This work was partially supported  by RFBR grant No.02-02-16957.

\end{document}